\documentclass[a4paper,11pt]{article}

\usepackage{epsfig}
\usepackage{amsmath,amssymb,amsfonts}
\usepackage{natbib}
\usepackage{bm}
\usepackage[svgnames]{xcolor}

\usepackage{mathptmx}

\newcommand{\sfrac}[2]{\mbox{$\frac{#1}{#2}$}}

\begin{document}
%%%%%%%%%%%%%%%%%%%%%%%%%%%%%%%%%%%%%%%%%%%%%%%%%%%%%%%%%%%%%%%%%%%%%%%%%%%%%
\title{Sampling on the sphere from $f(x)\propto x^TAx$}
\author{Richard Arnold\\ 
        School of Mathematics and Statistics\\
        Victoria University of Wellington\\
        Wellington, New Zealand\\ 
        {\em richard.arnold@vuw.ac.nz}} 
\date{\today} 
\maketitle
%%%%%%%%%%%%%%%%%%%%%%%%%%%%%%%%%%%%%%%%%%%%%%%%%%%%%%%%%%%%%%%%%%%%%%%%%%%%%
\section{Preliminaries}

We wish to sample on the surface of the sphere $S^{p-1} = \{x:
x\in\mathbb{R}^p, \ x^Tx=1\}$ from an axial distribution with density
proportional to $x^TAx$.

We assume that the matrix $A$ is symmetric, positive definite and can be
diagonalised
\[
   A=R\Lambda R^T
\]
for orthogonal matrix $R$ and diagonal
$\Lambda=\text{diag}(\lambda_1,\ldots,\lambda_p)$ with $\lambda_j>0$ for all
$j=1,\ldots,p$.

Using the notation from \citep{Arnold.Jupp:2013} we can construct the density
from an axial cardoid distribution
\begin{eqnarray}
   \nonumber
   f(x|\kappa,A) &=& 1 + \kappa\langle A, t([U])\rangle\\
   \nonumber
                 &=& 1 + \kappa\text{tr}\left(A(xx^T-I/p)\right)\\
   \nonumber
                 &=& 1 + \kappa\left(x^TAx-\text{tr}(A)/p\right)\\
   \label{eq:fulldensity}
                 &=& \frac{p}{\text{tr}(A)} x^TAx
\end{eqnarray}
where the last line follows from the requirement that
$1-\kappa\text{tr}(A)/p=0$ which ensures that the desired proportionality
holds.

In order to draw from this distribution we first take a draw $u$ from
\begin{equation}\label{eq:diagdensity}
   f(u|\kappa,\Lambda) = \frac{p}{\text{tr}(\Lambda)} u^T\Lambda u
                       = \frac{p}{\sum_{j=1}^p \lambda_j} \sum_{j=1}^p \lambda_ju_j^2
\end{equation}
and then compute $x=Ru$.  

%%%%%%%%%%%%%%%%%%%%%%%%%%%%%%%%%%%%%%%%%%%%%%%%%%%%%%%%%%%%%%%%%%%%%%%%%%%%%
\section{Construction}

The $p-$dimensional vector $u$ has $p-1$ degrees of freedom which we
parameterise as $(t_1,\ldots,t_{p-2},\phi_{p-1})$ with $t_j\in[-1,1]$ for
$j=1,\ldots,p-2$ and $\phi_{p-1}\in[0,2\pi)$ \citep[see,
    e.g.][Ch. 9]{Mardia_Jupp:2000}.

We then construct the vector $u$ as follows:
\begin{equation}\label{eq:construct.u}
   u = \begin{bmatrix} u_1\\ u_2\\ \vdots\\ u_j\\ \vdots\\ 
                       u_{p-2}\\ u_{p-1}\\ u_{p}
       \end{bmatrix}
     = \begin{bmatrix} t_1\\ (1-t_1^2)^{1/2} t_2\\ \vdots\\
                       \bigl[\prod_{\ell=1}^{j-1}(1-t_\ell^2)^{1/2}\bigr] t_j\\ \vdots\\
                       \bigl[\prod_{\ell=1}^{p-3}(1-t_\ell^2)^{1/2}\bigr] t_{p-2}\\ 
                       \bigl[\prod_{\ell=1}^{p-2}(1-t_\ell^2)^{1/2}\bigr] \cos\phi_{p-1}\\ 
                       \bigl[\prod_{\ell=1}^{p-2}(1-t_\ell^2)^{1/2}\bigr] \sin\phi_{p-1}
       \end{bmatrix}
     = \begin{bmatrix} t_1\\ b_2^{1/2} t_2\\ \vdots\\
                       b_j^{1/2} t_j\\ \vdots\\
                       b_{p-2}^{1/2} t_{p-2}\\ 
                       b_{p-1}^{1/2} \cos\phi_{p-1}\\ 
                       b_{p-1}^{1/2} \sin\phi_{p-1}
       \end{bmatrix}
\end{equation}
Here we have defined the symbols
\begin{eqnarray}
  b_1 &=& 1 \\
  b_j(t_1,\ldots,t_{j-1}) &=& \prod_{\ell=1}^{j-1}(1-t_\ell^2)
     \qquad\text{for $j=2,\ldots,p-1$}\\
  \nonumber
  &=& b_{j-1}(t_1,\ldots,t_{j-2}) (1-t_{j-1})^2
\end{eqnarray}
and we further define
\begin{eqnarray}
  a_1 &=& 0 \\
  \nonumber
  a_j(t_1,\ldots,t_{j-1}) &=& \sum_{k=1}^{j-1}\lambda_k t_k^2
                             \prod_{\ell=1}^{k-1}(1-t_\ell^2) \\
                         &=& \sum_{k=1}^{j-1}\lambda_k b_k t_k^2
     \qquad\text{for $j=2,\ldots,p-1$}\\
  \nonumber
  &=& a_{j-1}(t_1,\ldots,t_{j-2}) + \lambda_{j-1}b_{j-1}t_{j-1}^2
\end{eqnarray}
Under this parameterisation the uniform distribution on the sphere has the form
\begin{eqnarray}
  f_\text{Uniform}(t_1,\ldots,t_{p-2},\phi_{p-1})
  &=& 
  \frac{1}{2\pi} \prod_{j=1}^{p-2} B(\sfrac12, \sfrac{p-j}{2})^{-1}
                 (1-t_j^2)^{(p-j-2)/2}\\
  \nonumber
  &=&
  \frac{1}{2\pi} g_{p-2}(t_1,\ldots, t_{p-2})
\end{eqnarray}
where $B(a,b)$ is the Beta function $\Gamma(a)\Gamma(b)/\Gamma(a+b)$ and
we define
\begin{eqnarray}
  g_k(t_1,\ldots, t_k) &=& 
                 \prod_{j=1}^k B(\sfrac12, \sfrac{p-j}{2})^{-1}
                 (1-t_j^2)^{(p-j-2)/2}
  \qquad\text{for $k=1,\ldots,p-2$}\\
  \nonumber
  &=& g_{k-1}(t_1,\ldots, t_{k-1}) 
                 B(\sfrac12, \sfrac{p-k}{2})^{-1}
                 (1-t_k^2)^{(p-j-2)/2}
\end{eqnarray}
If the density is given by \eqref{eq:diagdensity} then with respect to our
chosen parameterisation of $u$ we have
\begin{eqnarray}
  \nonumber
  \lefteqn{f(t_1,\ldots,t_{p-2},\phi_{p-1})} &&\\
  &=& 
  \frac{1}{2\pi} g_{p-2}(t_1,\ldots, t_{p-2}) \frac{p}{\sum_{j=1}^p\lambda_j}
      \sum_{j=1}^p \lambda_j x_j^2\\
  \nonumber
  &=& 
  \frac{1}{2\pi} g_{p-2}(t_1,\ldots, t_{p-2}) \frac{p}{\sum_{j=1}^p\lambda_j}
      \left(\sum_{j=1}^{p-2} \lambda_j b_jt_j^2
            + 
            \lambda_{p-1} b_{p-1}\cos^2\phi_{p-1}
            + 
            \lambda_p b_{p-1}\sin^2\phi_{p-1}
      \right)\\
  \label{eq:joint.pdf}
  &=& 
  \frac{1}{2\pi} g_{p-2}(t_1,\ldots, t_{p-2}) \frac{p}{\sum_{j=1}^p\lambda_j}
      \left(a_{p-1}
            + 
            \lambda_{p-1} b_{p-1}\cos^2\phi_{p-1}
            + 
            \lambda_p b_{p-1}\sin^2\phi_{p-1}
      \right)
\end{eqnarray}
Integrating with respect to $\phi_{p-1}\in[0,2\pi)$ we obtain the marginal
  distribution over $(t_1,\ldots,t_{p-2})$:
\begin{equation}
  f(t_1,\ldots,t_{p-2}) = 
  g_{p-2}(t_1,\ldots, t_{p-2}) \frac{p}{\sum_{k=1}^p\lambda_k}
      \left(a_{p-1}
            + 
            \sfrac12 b_{p-1}(\lambda_{p-1}+\lambda_p)
      \right)
\end{equation}
Further integrations over $t_j\in[-1,+1]$ lead to the general expression
\begin{equation}
  f(t_1,\ldots,t_j) = 
  g_j(t_1,\ldots, t_j) \frac{p}{\sum_{k=1}^p\lambda_k}
      \left(a_{j+1}
            + 
            \frac{b_{j+1}}{p-j} \left(\sum_{k=j+1}^p \lambda_k\right)
      \right)
\end{equation}
which holds for $j=1,\ldots,p-2$.  When $j=1$ we ultimately have 
\begin{eqnarray}
  \nonumber
  f(t_1) 
  &=& g_1(t_1) \frac{p}{\sum_{k=1}^p\lambda_k}
      \left(a_2
            + 
            \frac{b_2}{p-1} \left(\sum_{k=2}^p \lambda_k\right)
      \right)\\
  \nonumber
  &=& g_1(t_1) \frac{p}{\sum_{j=1}^p\lambda_j}
      \left(\lambda_1t_1^2 + 
            \sfrac{1}{p-1}
            \left(\sum_{k=2}^p \lambda_k\right) (1-t_1^2)\right)\\
  &=& \frac{\lambda_1}{\sum_{j=1}^p\lambda_j}
      B(\sfrac32,\sfrac{p-1}{2}) t_1^2 (1-t_1^2)^{(p-3)/2}
     +\frac{\sum_{k=2}^p\lambda_k}{\sum_{k=1}^p\lambda_k}
      B(\sfrac12,\sfrac{p+1}{2}) (1-t_1^2)^{(p-1)/2}
\end{eqnarray}
This latter distribution is a mixture of two Beta densities in $t_1^2$.  To
simulate from it we proceed as follows:
\begin{enumerate}
\item Compute the mixture probabilities
   \begin{equation}
   \pi_1 = \frac{\lambda_1}{\sum_{j=1}^p\lambda_j}
   \qquad\text{and}\qquad
   \pi_2 = \frac{\sum_{k=2}^p\lambda_k}{\sum_{k=1}^p\lambda_k}
   \end{equation}
\item Draw $X_1\sim\text{Categorical}((1,2); (\pi_1,\pi_2))$
\item If $X_1=1$ draw 
   \[
       B^\ast_1 \sim \text{Beta}(\sfrac32,\sfrac{p-1}{2})
   \]
   otherwise if $X_1=2$ draw
   \[
       B^\ast_1 \sim \text{Beta}(\sfrac12,\sfrac{p+1}{2})
   \]
\item Draw $D_1\sim\text{Categorical}((-1,+1); (\sfrac12,\sfrac12))$
\item Set $t_1=D_1\sqrt{B^\ast_1}$.
\end{enumerate}
To simulate $t_2,\ldots,t_{p-2}$ we draw each $t_j$ conditional on
$t_1,\ldots,t_{j-1}$ by constructing the conditional distributions
\begin{eqnarray}
   \nonumber
   \lefteqn{f(t_j|t_1,\ldots,t_{j-1})} &&\\
   \nonumber
   &=& \frac{f(t_1,\ldots,t_{j-1},t_j)}{f(t_1,\ldots,t_{j-1})}\\
   \nonumber
   &=& \frac{g_j(t_1,\ldots, t_j)}{g_{j-1}(t_1,\ldots, t_{j-1})}
       \frac{
            a_{j+1}
            + 
            \frac{b_{j+1}}{p-j} \left(\sum_{k=j+1}^p \lambda_k\right)
        }{
            a_{j}
            + 
            \frac{b_{j}}{p-j+1} \left(\sum_{k=j}^p \lambda_k\right)
        }\\
   &=& \left[\frac{(1-t_j^2)^{(p+j-2)/2}}{B(\sfrac12,\sfrac{p-j}{2})}\right]
       \left[
       \frac{a_j + \lambda_jb_jt_j^2 
                 + \sfrac{1}{p-j}\left(\sum_{k=j+1}^p\lambda_k\right)
                   b_j(1-t_j^2)
             }{
             a_j + \sfrac{1}{p-j+1}\left(\sum_{k=j}^p\lambda_k\right)
                   b_j
            }\right]
\end{eqnarray}
This is a mixture of three Beta densities in $t_j^2$.  The mixing proportions
are
\begin{eqnarray} 
  \nonumber
  \pi_{j1} &=& \frac{a_j(p-j+1)}{a_j(p-s+1) + b_j(\sum_{k=j}^p\lambda_k)}\\
  \label{eq:tj.mixing.proportions}
  \pi_{j2} &=& \frac{b_j\lambda_j}{a_j(p-s+1) + b_j(\sum_{k=j}^p\lambda_k)}\\
  \nonumber
  \pi_{j3} &=& \frac{b_j(\sum_{k=j+1}^p\lambda_k)}{a_j(p-s+1) + b_j(\sum_{k=j}^p\lambda_k)}
\end{eqnarray}
corresponding to the three component densities 
$\text{Beta}(\sfrac12,\sfrac{p-j}{2})$, 
$\text{Beta}(\sfrac32,\sfrac{p-j}{2})$ and
$\text{Beta}(\sfrac12,\sfrac{p-j+2}{2})$ respectively. 

Thus to simulate $t_j|t_1,\ldots,t_{j-1}$
\begin{enumerate}
\item Compute the mixture probabilities $\pi_{j1}$, $\pi_{j2}$ and $\pi_{j3}$ using
   \eqref{eq:tj.mixing.proportions}
\item Draw $X_j\sim\text{Categorical}((1,2,3); (\pi_{j1},\pi_{j2},\pi_{j3}))$
\item If $X_j=1$ draw 
   \[
       B^\ast_j \sim \text{Beta}(\sfrac12,\sfrac{p-j}{2})
   \]
   otherwise if $X_j=2$ draw 
   \[
       B^\ast_j \sim \text{Beta}(\sfrac32,\sfrac{p-j}{2})
   \]
   otherwise if $X_j=3$ draw
   \[
       B^\ast_j \sim \text{Beta}(\sfrac12,\sfrac{p-j+2}{2})
   \]
\item Draw $D_j\sim\text{Categorical}((-1,+1); (\sfrac12,\sfrac12))$
\item Set $t_j=D_j\sqrt{B^\ast_j}$.
\end{enumerate}
We carry out this procedure for $j=2,\ldots,p-2$.  Note that the procedure is
also defined for $j=1$, in which case $\pi_1=0$ and we have the same two
component mixture derived earlier.

The final step is to draw the angle $\phi_{p-1}\in[0,2\pi)$.  Its conditional
  distribution is given by 
\begin{eqnarray}
   \nonumber
   \lefteqn{f(\phi_{p-1}|t_1,\ldots,t_{p-2})} &&\\
   \nonumber
   &=& \frac{f(t_1,\ldots,t_{p-2},\phi_{p-1})}{f(t_1,\ldots,t_{p-2})}\\
   \nonumber
   &=&
   \frac{1}{2\pi}
   \frac{
         a_{p-1}
            + 
            \lambda_{p-1} b_{p-1}\cos^2\phi_{p-1}
            + 
            \lambda_p b_{p-1}\sin^2\phi_{p-1}
         }{
         a_{p-1}
            + 
            \sfrac12 b_{p-1}(\lambda_{p-1}+\lambda_p)
         }\\
   &=&
   \frac{1}{2\pi}
   \left[1 + \frac{
            \sfrac{1}{2}b_{p-1}(\lambda_{p-1} -\lambda_p)\cos 2\phi_{p-1}
         }{
         a_{p-1}
            + 
            \sfrac12 b_{p-1}(\lambda_{p-1}+\lambda_p)
         }\right]
\end{eqnarray}
This density has four-fold symmetry: $\pm\phi$ and$\pi\pm\phi$ are all
equivalent.  To draw a value for $\phi_{p-1}$ we first draw
$U_{p-1}\sim\text{Uniform}(0,1)$ and then solve the equation
\begin{eqnarray}
   \sfrac14 U_{p-1} &=& 
         \int_0^{\phi^\ast} f(\phi'|t_1,\ldots,t_{p-2})\,{\rm d}\phi'\\
   &=&
   \frac{1}{2\pi}
   \left[\phi^\ast + \frac{
            \sfrac{1}{4}b_{p-1}(\lambda_{p-1} -\lambda_p)\sin 2\phi^\ast
         }{
         a_{p-1}
            + 
            \sfrac12 b_{p-1}(\lambda_{p-1}+\lambda_p)
         }\right]
\end{eqnarray} 
in the interval $\phi^\ast\in[0,\sfrac{\pi}{2}]$.  We finally set $\phi_{p-1}$
to one of the four values $\{\pm\phi^\ast, \pi\pm\phi^\ast\}$ with equal
probability.

Once we have a full draw of the values $(t_1,\ldots,t_{p-2},\phi_{p-1})$ we 
use \eqref{eq:construct.u} to construct the vector $u$, and finally set $x=Ru$
as the random draw from $f(x)\propto x^TAx$.  

%%%%%%%%%%%%%%%%%%%%%%%%%%%%%%%%%%%%%%%%%%%%%%%%%%%%%%%%%%%%%%%%%%%%%%%%%%%%%
\section{R code}

The following R function can be used to take $n={\tt n}$ draws from the density
\eqref{eq:fulldensity} where $A={\tt amat}$ is a $p\times p$ square, positive
defnite matrix.  The function returns an $n\times p$ matrix, the rows of which
are the $n$ random vectors. 

\begin{quote}
\begin{small}
\begin{verbatim}
rxax <- function(n,amat) {
    # Eigen decomposition
    ee <- eigen(amat)
    dvec <- ee$values
    rmat <- ee$vectors
    csum <- rev(cumsum(rev(dvec)))
    p <- nrow(amat)

    # Generate vectors in rotated frame where amat is diagonal
    umat <- sapply(1:n,
                   function(i) {
                      avec <- rep(NA,p-1)
                      bvec <- rep(NA,p-1)
                      tvec <- rep(NA,p-2)
                      # generate t[1]...t[p-2]
                      avec[1] <- 0
                      bvec[1] <- 1
                      for(j in 1:(p-2)) {
                          pivec <- c( avec[j]*(p-j+1), 
                                      bvec[j]*dvec[j],
                                      bvec[j]*csum[j+1] )
                          pivec <- pivec/sum(pivec)
                          i <- sample(1:3, size=1, prob=pivec)
                          b <- rbeta(1,
                                     (c(1,3,1)/2)[i],
                                     (c(p-j,p-j,p-j+2)/2)[i])
                          d <- sample(c(-1,1), size=1)
                          tvec[j] <- d*b
                          bvec[j+1] <- bvec[j]*(1-tvec[j]^2)
                          avec[j+1] <- avec[j] + dvec[j]*bvec[j]*tvec[j]^2
                      }
                      # generate phi[p-1]
                      ustar <- runif(1)
                      ffunc <- function(phi, uval, a, b, d1, d2) {
                          c1 <- (b/4)*(d1-d2)
                          c2 <- a + 0.5*b*(d1+d2)
                          (1/(2*pi)*( phi + c1*sin(2*phi)/c2 ) - uval/4)
                      }
                      # generate angle in (0,pi/2)
                      phistar <- uniroot(ffunc, c(0,pi/2),
                                         uval=ustar, a=avec[p-1], b=bvec[p-1],
                                         d1=dvec[p-1], d2=dvec[p])$root
                      # randomise over fourfold symmetry
                      phistar <- (sample(c(0,1),size=1)*pi
                                 +sample(c(-1,+1),size=1)*phistar)
                      
                      # construct uvec
                      uvec <- c( sqrt(bvec)[1:(p-2)]*tvec,
                                 sqrt(bvec[p-1])*c(cos(phistar), sin(phistar)) )
                      })
    # rotate back to natural frame
    xmat <- t(rmat%*%umat)
    return(xmat)
}
\end{verbatim}
\end{small}
\end{quote}

An example use of the function is as follows: we construct a suitable symmetric
$3\times 3$ matrix $A$ at random and then take 10 draws from it.
\begin{quote}
\begin{small}
\begin{verbatim}
amat <- array(rnorm(9),dim=c(3,3))
amat <- t(amat)%*%amat
xmat <- rxax(10, amat)
xmat

             [,1]        [,2]       [,3]
 [1,]  0.98674959  0.01738693 -0.1613163
 [2,] -0.38065170  0.23323003  0.8948229
 [3,]  0.24385388 -0.60136593  0.7608510
 [4,] -0.44503351 -0.66007152  0.6051865
 [5,] -0.85847885 -0.04198661 -0.5111274
 [6,]  0.33440170  0.04373660  0.9414152
 [7,] -0.09637544 -0.11457369 -0.9887288
 [8,]  0.27959151 -0.70197708  0.6550242
 [9,]  0.61130650 -0.18851719  0.7686128
[10,] -0.38150181 -0.76106892  0.5246241
\end{verbatim}
\end{small}
\end{quote}

% Citations that may be relevant
%\citet{Ley.etal:2013}
%%%%%%%%%%%%%%%%%%%%%%%%%%%%%%%%%%%%%%%%%%%%%%%%%%%%%%%%%%%%%%%%%%%%%%%%%%%%%
\bibliographystyle{agsm}
\bibliography{directional}
%%%%%%%%%%%%%%%%%%%%%%%%%%%%%%%%%%%%%%%%%%%%%%%%%%%%%%%%%%%%%%%%%%%%%%%%%%%%%
\end{document}